# Low-Frequency Electronic Noise in β-(Al$_x$Ga$_{1-x}$)$_2$O$_3$ Schottky Barrier Diodes

Subhajit Ghosh[1], Dinusha Herath Mudiyanselage[2], Sergey Rumyantsev[3], Yuji Zhao[4], Houqiang Fu[2], Stephen Goodnick[2], Robert Nemanich[5] and Alexander A. Balandin[1],[×]

[1]Department of Electrical and Computer Engineering, University of California, Riverside, California 92521 USA

[2]School of Electrical, Computer, and Energy Engineering, Arizona State University, Tempe, Arizona 85287 USA

[3]CENTERA Laboratories, Institute of High-Pressure Physics, Polish Academy of Sciences, Warsaw 01-142 Poland

[4]Department of Electrical and Computer Engineering, Rice University, Houston, Texas 77005, USA

[5]Department of Physics, Arizona State University, Tempe, Arizona 85281 USA

**Abstract**

We report on the low-frequency electronic noise in β-(Al$_x$Ga$_{1-x}$)$_2$O$_3$ Schottky barrier diodes. The noise spectral density reveals 1/$f$ dependence, characteristic of the flicker noise, with superimposed Lorentzian bulges at the intermediate current levels ($f$ is the frequency). The normalized noise spectral density in such diodes was determined to be on the order of 10$^{-12}$ cm$^2$/Hz ($f$=10 Hz) at 1 A/cm$^2$ current density. At the intermediate current regime, we observed the random telegraph signal noise, correlated with the appearance of Lorentzian bulges in the noise spectrum. The random telegraph signal noise was attributed to the defects near the Schottky barrier. The defects can affect the local electric field and the potential barrier, and correspondingly, impact the electric current. The obtained results help to understand noise in Schottky barrier diodes made of ultra-wide-band-gap semiconductors and can be used for the material and device quality assessment.

**Keywords:** 1/$f$ noise; flicker noise; random telegraph signal noise; Schottky diode; reliability

[×] Corresponding author (AAB); E-mail: balandin@ece.ucr.edu ;   https://balandingroup.ucr.edu/

Ultra-wide bandgap (UWBG) semiconductors are attracting increasing attention owing to the industry's need for high-power density electronics[1–5]. While GaN and SiC technologies are already well established, diamond and β-Ga$_2$O$_3$ are other promising materials for power switches and other electronic devices. High-quality epitaxial β-Ga$_2$O$_3$ films can be grown by metal-organic-chemical-vapor deposition (MOCVD) on native substrates with controlled doping levels[6–10]. Merging Al$_2$O$_3$ with β-Ga$_2$O$_3$ can result in an alloy with increased bandgap and improved performance in comparison to β-Ga$_2$O$_3$[11]. The effect of the alloy composition on the barrier height has been investigated in details[12]. Most recently, a number of β-(Al$_x$Ga$_{1-x}$)$_2$O$_3$ electronic and optoelectronic devices have been demonstrated[13–23]. The vertical Schottky barrier diodes (SBDs) based on β-(Al$_x$Ga$_{1-x}$)$_2$O$_3$ appear to be among the most promising. Naturally, as with any novel technology, one has to spend significant efforts in improving the quality of the material and understanding the effects of defects on the device's performance.

Measurements of low-frequency noise provide valuable information on the material quality and device reliability[24–36]. The low-frequency noise includes the 1/$f$ noise, also known as flicker noise, and generation-recombination (G-R) noise with a Lorentzian-type spectrum ($f$ is the frequency). Both flicker and G-R noise are typically associated with defects that act as trapping centers for the charge carriers. The noise level can be used as a metric to assess the maturity of the device technology. For example, the noise spectral density in GaN field-effect transistors has shown a decrease of over six orders of magnitude as the technology improved[37,38]. In addition, low-frequency noise can be used as an indicator of device damage since noise is sensitive to defects, electromigration, and leakage currents. We have previously reported on noise in high-current GaN and diamond PIN diodes[39,40]. It would be highly desirable for the UWBG technologies to measure noise in β-(Al$_x$Ga$_{1-x}$)$_2$O$_3$ SBDs and compare its level and characteristics with those in GaN PIN diodes and diamond PIN and SBD devices. There is limited data available for 1/$f$ noise in β-Ga$_2$O$_3$ thin films and devices[41–43]. We are not aware of any reports of noise in β-(Al$_x$Ga$_{1-x}$)$_2$O$_3$ materials or devices.

In this Letter, we present the results of the investigation of noise in β-(Al$_x$Ga$_{1-x}$)$_2$O$_3$ ($x$ = 0.21) SBDs. For this study, we selected devices designed specifically for applications such as high-current switches in smart electricity grids and related electronics. The device fabrication steps included molecular beam epitaxy (MBE) growth of β-(Al$_x$Ga$_{1-x}$)$_2$O$_3$ epilayers on the edge-defined film-fed grown (010) β-Ga$_2$O$_3$ substrate. The details of the growth of β-(Al$_x$Ga$_{1-x}$)$_2$O$_3$ epilayer were reported by some of us elsewhere[44]. The β-Ga$_2$O$_3$ substrate was heavily doped with Sn to aid the formation of the back Ohmic contact. The layer of β-(Al$_x$Ga$_{1-x}$)$_2$O$_3$ was $n$-type doped with Si. The specific design and the doping levels are indicated in Figure 1 (a). Photolithography, inductively coupled plasma reactive ion etching (ICP-RIE), and electron beam (e-beam) evaporation were used to fabricate the devices. A cathode made of Ti/Au (20/130) nm was deposited at the back side of the β-Ga$_2$O$_3$ substrate followed by 500 °C rapid thermal annealing (RTA) in N$_2$ environment. After that step, the front contact vias were formed using photolithography and development to form the Schottky contact. Finally, Pt/Ti/Au (20/10/120) nm anode, *i.e.,* the Schottky contact, was deposited by e-beam evaporation; liftoff was utilized to isolate individual devices. For the current–voltage (I–V) measurements, the substrate containing the β-(Al$_x$Ga$_{1-x}$)$_2$O$_3$ diodes was placed inside the chamber of the Lakeshore TTPX probe station. The chamber pressure was lowered to 10$^{-5}$ Torr. The DC I–V measurements were conducted in the 2-terminal configuration using a semiconductor analyzer (Agilent B1500). Figure 1 (b) shows forward-bias I–V characteristics of a representative 100-μm diameter diode at room temperature (RT). The device reveals an ideality factor, $n$, of ~1.65 in the low-bias region. The effective Schottky barrier for this device, calculated from the thermionic emission model, is ~0.9 eV at RT. The ideality factor for all studied devices varied between $n$ = 1.2 and $n$ = 1.8.

The noise measurements were conducted following the standard protocol[39,40]. The noise system consisted of the device under test connected in series with a load resistor and a DC biasing battery. A potentiometer was used to control the voltage drop across the voltage divider circuit. During the noise measurements, the voltage fluctuations were amplified with the low-noise voltage preamplifier (SR-560). The amplified voltage signal was converted to its corresponding frequency spectrum using a dynamic signal analyzer (Photon+). For the noise data analysis, the measured voltage noise spectral density, $S_V$, was converted to its equivalent short circuit current spectral density, $S_I$. The details of the noise measurements and data analysis were reported by some of us in the context of other materials and devices[39,40,45–48]. The noise spectra, $S_I$, of β-(Al$_x$Ga$_{1-x}$)$_2$O$_3$ SBD, with the I-Vs shown in Figure 1 (b), are presented in Figure 2 (a). The data are shown for the low and intermediate current regimes, for the current in the range from 2×10$^{-9}$ A to 3×10$^{-6}$ A. In Figure 2 (a), we show the raw data, not normalized by the current, to demonstrate the regions with 1/$f$ spectrum and Lorentzian bulges more explicitly. Figure 2

(b) presents the noise spectral density multiplied by the frequency, $S_I \times f$, to eliminate the $1/f$ flicker noise background. One can see that the superimposed Lorentzian bulges are pronounced at the intermediate current regime.

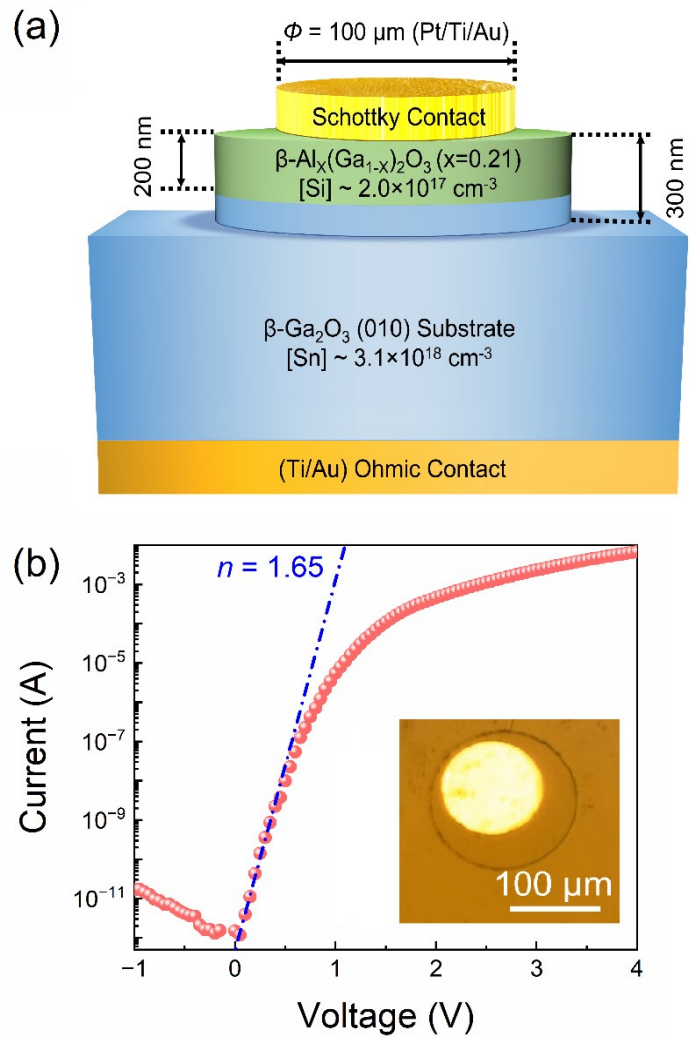

**Figure 1**: (a) Schematic of the vertical β-(Al$_x$Ga$_{1-x}$)$_2$O$_3$ ($x$ = 0.21) SBDs. (b) I–V characteristics of the diode plotted in the semi-log scale. The dashed line identifies $n$. The inset shows the optical microscopy image of several diodes.

The normalized noise current spectral density, $S_I/I^2$ at the intermediate and high current regimes, with the current varying from $5\times10^{-6}$ A to $3\times10^{-3}$ A, is presented in Figure 2 (c). In Figure 2 (d) we show the corresponding normalized noise spectral density multiplied by the frequency, $S_I/I^2 \times f$. The conclusion from Figure 2 (a-d) is that the noise in β-(Al$_x$Ga$_{1-x}$)$_2$O$_3$

SBD is mostly of the 1/$f$ flicker type with the Lorentzian budges superimposed on the 1/$f$ background in the intermediate forward current regime. The Lorentzian spectral features are clearly seen at the current levels from $3\times10^{-7}$ A to $1\times10^{-5}$ A, a region that corresponds to the onset of the I–V bending in the I–Vs (see Figure 1 (b)). The Lorentzians observed in the low-frequency noise spectra can be associated with the generation-recombination (G-R) noise mechanism. A detailed discussion of the origin of the Lorentzian bulges is provided below.

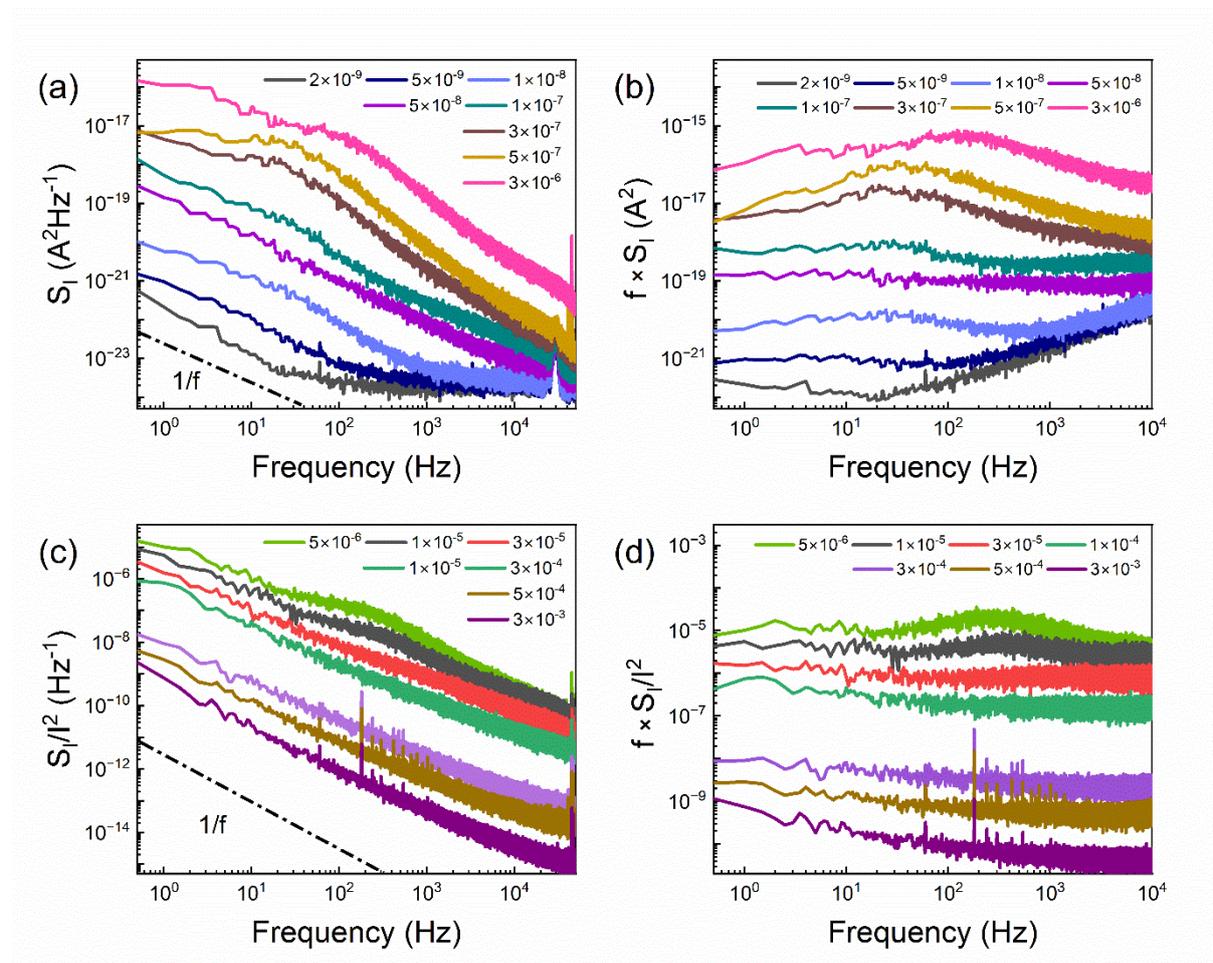

**Figure 2**: (a) The noise current spectral density, $S_I$, as a function of frequency, $f$ for several values of the forward current in the low and intermediate currents (up to $I$ = 3 µA). (b) The current spectral density, multiplied by the frequency, $S_I \times f$, as a function of frequency, $f$, plotted for the current values till $I$ = 3 µA. (c) The normalized current spectral density, $S_I/I^2$, as a function of frequency, $f$, for several values of the forward current in the intermediate (from $I$ = 5 µA) and high current regimes. (d) The normalized current spectral density, multiplied by the frequency, $S_I/I^2 \times f$ vs frequency, $f$, dependence at higher forward currents (up to $I$ = 1 mA). The current values in the legends are indicated in amperes (A).

It is useful to compare the noise level at different current regimes and address the question "How noisy are β-(Al$_x$Ga$_{1-x}$)$_2$O$_3$ SBDs compared to high-current diodes made of other UWBG semiconductors?" Figure 3 (a) shows the current noise spectral density, $S_I$, measured at $f = 10$ Hz, as a function of the forward current. The $S_I$ vs $I$ follows a quadratic relationship in the lower current regime, i.e., $S_I \sim I^2$. The dependence experiences a change in the intermediate and high current regimes. The variations in the dependence can be related to the Lorentzian spectral features, which appear at the same current levels. In general, the G-R-type noise spectrum shape depends on the occupancy of the trap responsible for noise and the concentration of free charge carriers. Since the position of the Fermi level in the space-charge region of the SBD depends on the bias, the amplitude, and the characteristic frequency of the Lorentzian bulges in the noise also depend on current. This explains the deviation in the noise spectral density dependence on current from the conventional $S_I \sim I^2$ trend.

In Figure 3 (b), we show the noise spectral density normalized by the current and device area, $S_I/I^2 \times \Omega$, for three different UWBG device technologies: β-(Al$_x$Ga$_{1-x}$)$_2$O$_3$ SBDs, diamond PIN-SBD devices[39] and GaN PIN diodes[40]. One can see that in the entire current density range, β-(Al$_x$Ga$_{1-x}$)$_2$O$_3$ diodes have a noise level comparable to the diamond diodes. The noise level in the GaN diodes at any current is substantially lower than that in β-(Al$_x$Ga$_{1-x}$)$_2$O$_3$ and diamond diodes. The overall noise level has been used as a metric of the maturity of the technology, including material quality and device processing steps[26,39,40,49,50]. One can rationalize our results noting that β-(Al$_x$Ga$_{1-x}$)$_2$O$_3$ and diamond diodes are the newest and less mature technologies than GaN PINs.

Now we look closer to the noise behavior in the intermediate current regime, for the range from $I = 3 \times 10^{-7}$ A to $I = 1 \times 10^{-5}$ A, where the noise spectra show Lorentzian features. The Lorentzians can indicate the conventional G-R noise mechanism when one type of trap, with a specific time constant, has a substantially higher concentration than other traps and starts dominating the noise spectrum[24,39,40,51,52]. In this case, the Lorentzian associated with this certain time constant is superimposed on the $1/f$ envelope originating from all other traps with different time constants. If the Lorentzian features are accompanied by the random telegraph signal (RTS) noise in the time domain, they may indicate that just one or a few defect states act as trapping centers and produce a strong effect on noise while the other defects are absent. The latter is usually observed in nanoscale devices where just a few traps are present or at low temperatures where only a few traps are thermally activated. To understand the origin of the Lorentzian spectra in our devices we analyzed the time-domain noise response at six different current levels.

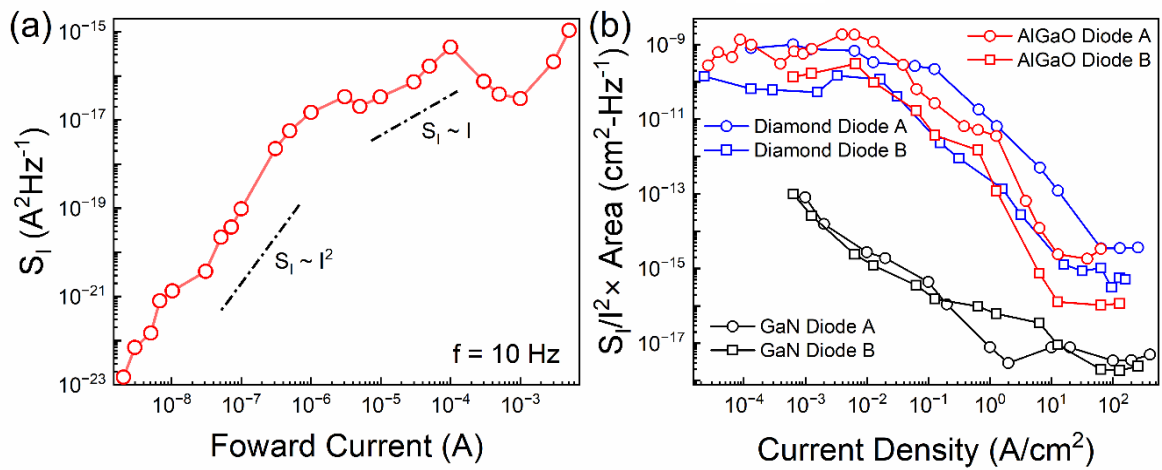

**Figure 3**: (a) The noise current spectral density, $S_I$, as a function of the forward current ($I$), at $f$ = 10 Hz, measured at room temperature. (b) The noise level ($S_I/I^2 \times \Omega$, where $\Omega$ is the device contact area) *vs* current density ($J$) of the vertical β-($Al_xGa_{1−x}$)$_2O_3$ SBDs as compared with the corresponding noise levels of the GaN PIN diodes and diamond diodes. The data points for GaN and diamond devices are from Refs.[39] and [40].

Figure 4 (a-f) shows the current fluctuations, Δ$I$, as the function of time, $t$. The current fluctuations reveal pulses with fixed amplitude but random width and duration. This time-domain response is characteristic of the RTS noise. Most commonly, RTS appears when the device has just one or a few fluctuators. The fluctuations between two states, *e.g.,* charged and non-charged trap, will appear as RTS in the time domain and as a Lorentzian in the spectrum. The examined SBDs have large dimensions and were tested at RT. The amplitudes of the current steps are in the nA range while the current is in the μA range. This type of RTS noise has been previously observed in SBDs and it was termed burst noise[53–57]. The dimensions of our devices and the amplitude of the RTS steps suggest that the RTS noise is likely due to a defect, or a few defects, near the Schottky barrier, *i.e.,* at the metal–semiconductor interface. The RTS noise in large devices at RT cannot be explained by a simple variation in the number of charge carriers *via* capture and emission by the trap, which results in electric current fluctuations. A trap at the Schottky barrier, which changes its state from neutral to charged and back, can cause a local variation in the electric field and affect the potential barrier. The changes in the height of the potential barrier can result in observable step changes in the current.

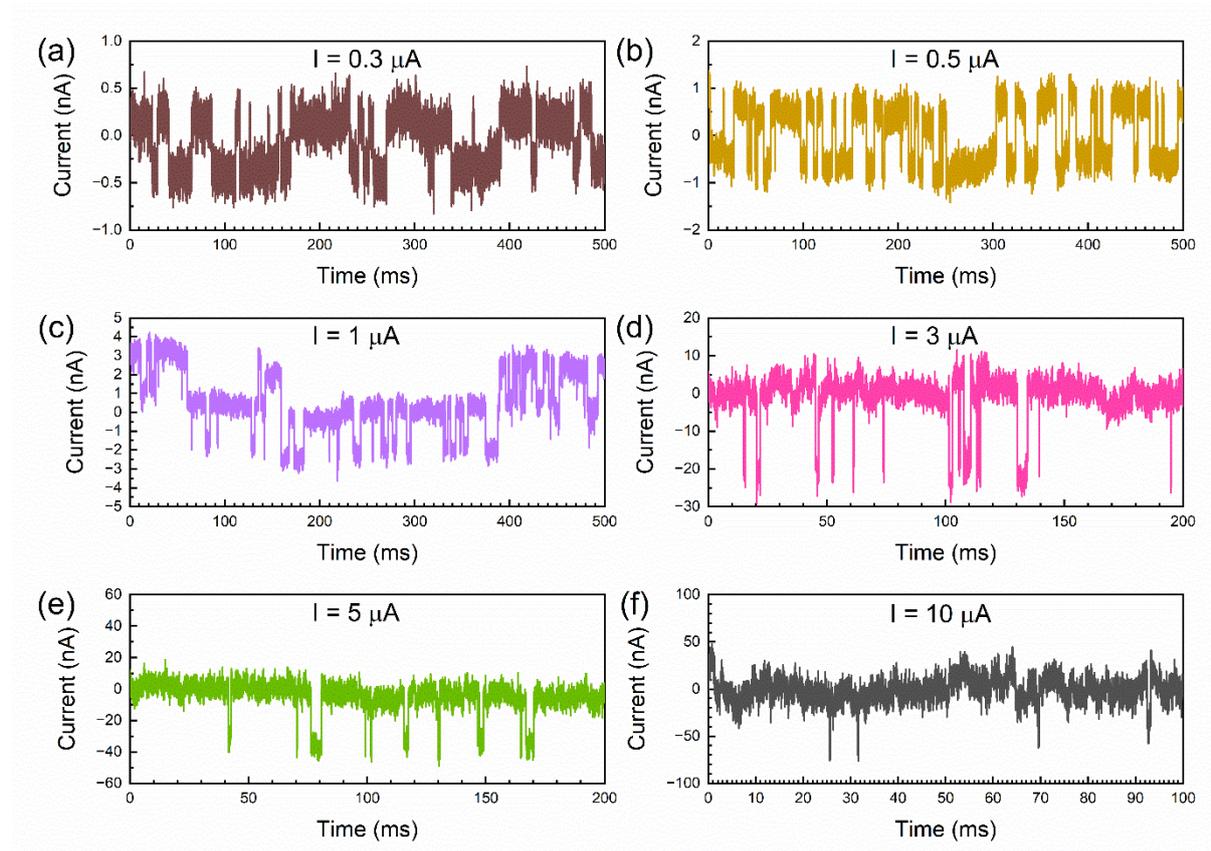

**Figure 4**: The time-domain current fluctuations shown for six intermediate current values of (a) $I$ = 0.3 µA, (b) $I$ = 0.5 µA, (c) $I$ = 1 µA, (d) $I$ = 3 µA, (e) $I$ = 5 µA, (f) $I$ = 10 µA. Note the appearance of the RTS noise.

In Figure 5 (a-d) we directly compare the time-domain current fluctuations with the corresponding frequency domain noise spectra at four different intermediate current levels. The detailed analysis of the noise spectral density plots shows the superposition of the Lorentzians of the RTS noise with the $1/f$ flicker noise. Using the linear and Lorentzian fitting we separate the RTS noise and the flicker noise envelope. The $S_I \sim 1/f^\zeta$ dependence shows the extracted $\zeta$ value in the range from 1.04 to 1.39. Since RTS noise is observed for the same current range as Lorentzian spectra we conclude that the Lorentzian features in the noise are of the RTS noise origin. The corner frequency $f_c$ of the Lorentzians moves towards higher frequencies with the increasing current. The $f_c$ is defined by the characteristic time $\tau=1/2\pi f_c$. This time constant depends of the characteristic emission and capture times and can be written as $\tau=\tau_c F$, where $\tau_c=(\sigma n v)^{-1}$, is characteristic capture time, $\sigma$ is the capture cross-section, $v$ is the thermal velocity, and $F$ is the occupancy function for the level responsible for noise[58]. With the increase of the bias voltage the height of the barrier decreases. As a result, the carrier concentration close to the barrier top increases, and the occupancy function of the levels in the same region decreases. Both these processes lead to $f_c$ increase as observed in the experiments. This supports our explanation that the RTS noise originates from one or a few defects at the metal–semiconductor interface, which affect the local electric field and potential barrier, thus producing a strong effect on the current. In this scenario, the main contribution to noise comes from a narrow region close to the barrier top where the concentration of carriers is exponentially small. Modulation of the barrier height, $\Delta\Phi$, due to the change in the trap charge state causes fluctuations in the current, which depends on the barrier highest, $\Phi$, exponentially. The nature of the defects acting as the trap levels at this point is not known. One of the likely possibilities for this material system can be alloy related defects[59].

Using time-domain RTS noise data, *e.g.* shown in Figure 4 and 5, we can plot the normalized amplitudes of the current steps for further analysis. The amplitudes of the current steps, $\Delta I_{RTS}/I$, for two representative devices are shown in Figure 6. Although the effect of the Coulombic charge on the barrier height depends on the exact location of the trap, the change in the barrier height can be roughly estimated as[57]:

$$\Delta\Phi = nkT \frac{\Delta I_{RTS}}{I}. \qquad (1)$$

This barrier height modulation should not depend significantly on the bias[57,60]. Within this scenario, we can divide the experimental points shown in Figure 6 into three groups with the weak dependences of $\Delta I_{RTS}/I$ on current, $I$, in each group. Assuming that each group of data points belongs to one trap we can distinguish two traps for device A and one trap for device B. Taking an average for the $\Delta I_{RTS}/I$ value, we can estimate the change in the barrier height $\Delta\Phi$ (shown in Figure 6 by the dashed lines). Since just one or a few traps, present within the whole area of the diode, cause the RTS noise, we should expect different trap parameters for different devices. The difference in $\Delta\Phi$ mainly originates from the trap position within the space charge region. The traps, which are closer to the metal–semiconductor interface, *i.e.*, closer to the barrier top, are characterized by higher $\Delta\Phi$. The values $\Delta\Phi$ extracted from the noise data are in the range from ~0.17 meV to ~0.45 meV (see Figure 6). These values are in the range below 1 meV, which is consistent with the values determined for the diodes made of wide-band-gap (WBG) materials such as SiC[57].

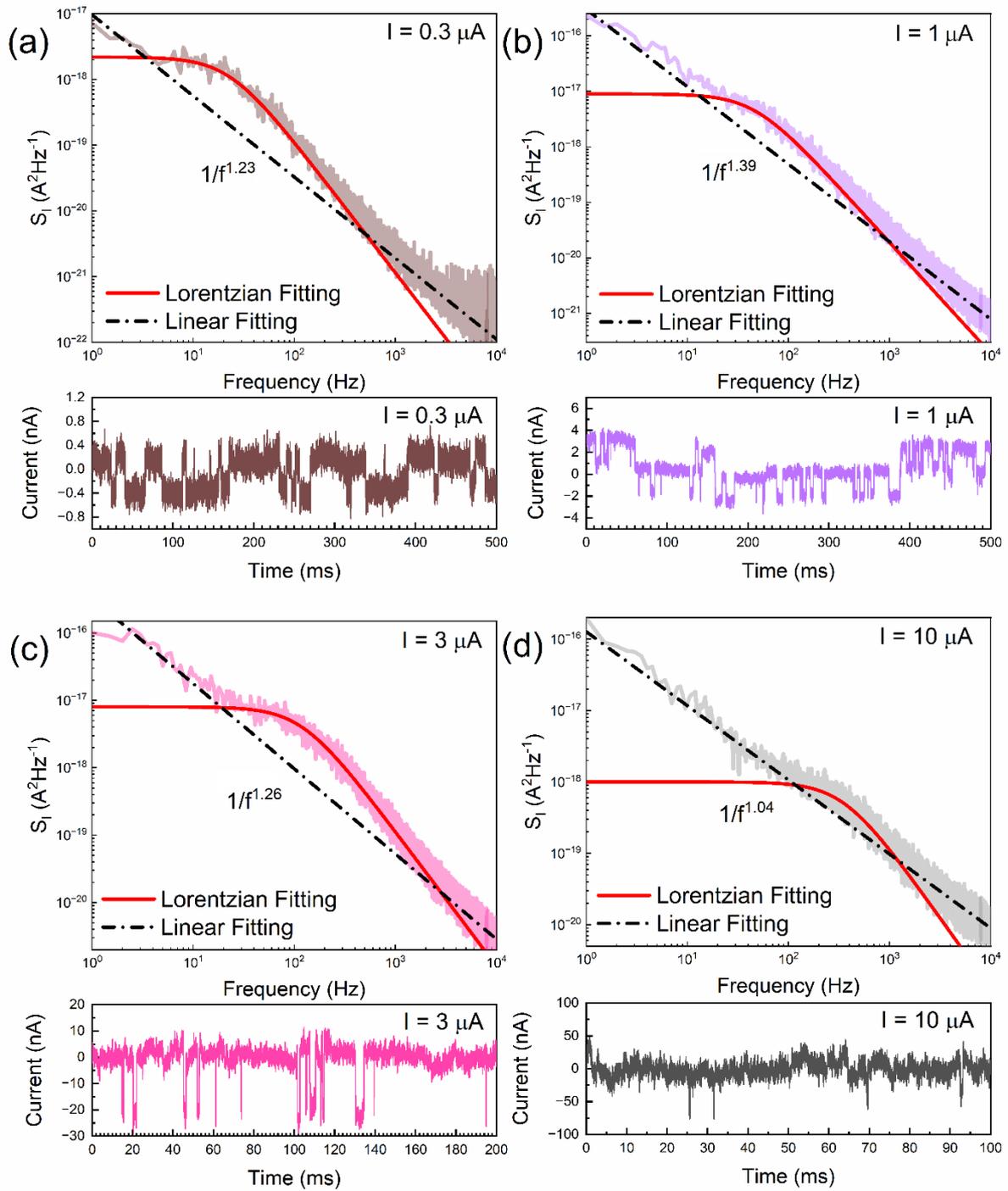

**Figure 5**: The frequency and time-domain noise response shown for four different intermediate current values of (a) $I$ = 0.3 µA, (b) $I$ = 1 µA, (c) $I$ = 3 µA, (d) $I$ = 10 µA.

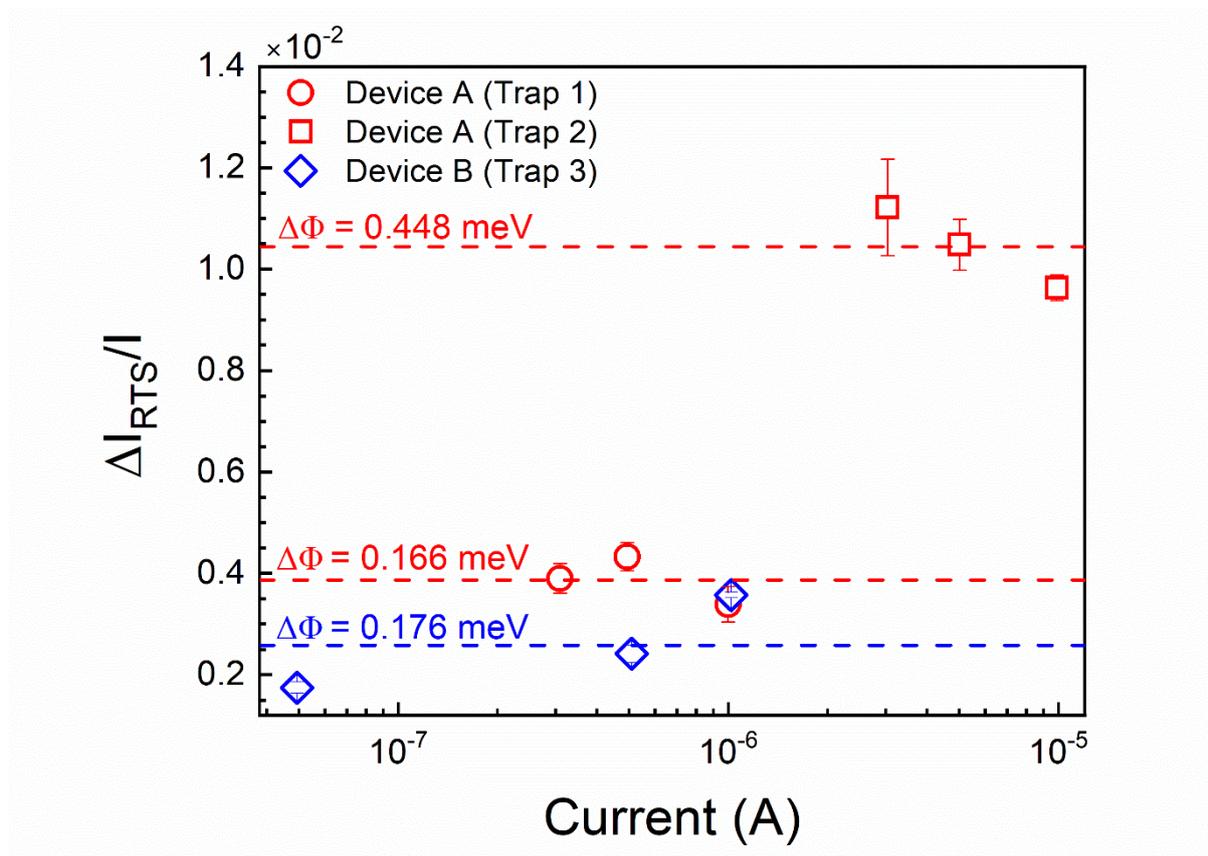

**Figure 6**: The amplitude of RTS noise normalized by the diode current, $\Delta I_{RTS}/I$, plotted as a function of the diode forward currents for two different diode devices. The dashed lines indicate the average values of $\Delta I_{RTS}/I$ for three different trap levels.

In summary, we reported on the low-frequency electronic noise in β-(Al$_x$Ga$_{1-x}$)$_2$O$_3$ SBDs. The noise spectral density reveals $1/f$ dependence, characteristic of the flicker noise, with superimposed Lorentzian bulges. At the intermediate current regime, we observed the RTS noise, correlated with the appearance of Lorentzian bulges in the noise spectrum. The RTS noise was attributed to the defects near the Schottky barrier. The defects can affect the local electric field and the potential barrier, and correspondingly, impact the electric current. The modulated barrier height $\Delta\Phi$ extracted from the noise data is in the range from ~0.17 meV to ~0.45 meV. The obtained results help to understand noise in Schottky barrier diodes made of UWBG semiconductors and can be used for the material and device quality assessment.

**Acknowledgments**

The work at UCR and ASU was supported by ULTRA, an Energy Frontier Research Center (EFRC) funded by the U.S. Department of Energy, Office of Science, Basic Energy Sciences under Award # DE-SC0021230. S.R. who contributed to noise data analysis acknowledges the support by CENTERA Laboratories in a frame of the International Research Agendas program for the Foundation for Polish Sciences co-financed by the European Union under the European Regional Development Fund (No. MAB/2018/9). S.G. and A.A.B thank Fariborz Kargar for useful discussions.

**Conflict of Interest**

The authors declare no conflict of interest.

**Author Contributions**

A.A.B. coordinated the project and led the data analysis, and manuscript preparation. S.G. measured I-Vs and noise characteristics, and contributed to the data analysis; D.H.M. fabricated Schottky barrier diodes and measured I-Vs; Y.Z. and H.F. supervised device fabrication; S.R., S.G., and R.N. contributed to the diode I-V and noise data analysis. All authors contributed to the manuscript preparation.

**The Data Availability Statement**

The data that support the findings of this study are available from the corresponding author upon reasonable request.